\documentclass[twocolumn]{aastex63}
\usepackage[utf8]{inputenc}
\usepackage{graphicx}  
\usepackage{float}  
\usepackage{subfigure}  
\usepackage{CJK}
\usepackage{epsfig}
\usepackage{amsmath}

\shortauthors{J.-T. Li et al.}
\shorttitle{eDIG-CHANGES II: NGC~3556}

\begin{document}
\begin{CJK*}{UTF8}{gbsn}

\title{eDIG-CHANGES II: Project Design and Initial Results on NGC~3556}

\author[0000-0001-6239-3821]{Jiang-Tao Li (李江涛)}
\affiliation{Purple Mountain Observatory, Chinese Academy of Sciences, 10 Yuanhua Road, Nanjing 210023, People's Republic of China}

\author[0000-0002-3286-5346]{Li-Yuan Lu (芦李源)}
\affiliation{Department of Astronomy, Xiamen University, 422 Siming South Road, Xiamen, Fujian, People's Republic of China}
\affiliation{Purple Mountain Observatory, Chinese Academy of Sciences, 10 Yuanhua Road, Nanjing 210023, People's Republic of China}

\author[0000-0002-2941-646X]{Zhijie Qu (屈稚杰)}
\affiliation{Department of Astronomy \& Astrophysics, The University of Chicago, Chicago, IL 60637, U.S.A.}

\author[0000-0002-8109-2642]{Robert A. Benjamin}
\affiliation{Department of Physics, University of Wisconsin - Whitewater, 800 West Main Street, Whitewater, WI 53190, USA}

\author[0000-0001-6276-9526]{Joel N. Bregman}
\affiliation{Department of Astronomy, University of Michigan, 311 West Hall, 1085 S. University Ave, Ann Arbor, MI, 48109-1107, U.S.A.}

\author[0000-0001-8206-5956]{Ralf-J\"{u}rgen Dettmar}
\affiliation{Ruhr University Bochum, Faculty of Physics and Astronomy, Astronomical Institute (AIRUB), 44780 Bochum, Germany}

\author{Jayanne English}
\affiliation{Department of Physics \& Astronomy, University of Manitoba, Winnipeg, Manitoba, R3T 2N2, Canada}

\author[0000-0002-2853-3808]{Taotao Fang (方陶陶)}
\affiliation{Department of Astronomy, Xiamen University, 422 Siming South Road, Xiamen, Fujian, People's Republic of China}

\author[0000-0003-0073-0903]{Judith A. Irwin}
\affiliation{Department of Physics, Engineering Physics \& Astronomy, Queens University, Kingston, Ontario, K7L 3N6, Canada}

\author{Yan Jiang (姜燕)}
\affiliation{Purple Mountain Observatory, Chinese Academy of Sciences, 10 Yuanhua Road, Nanjing 210023, People's Republic of China}
\affiliation{School of Astronomy and Space Sciences, University of Science and Technology of China, Hefei 230026, People's Republic of China}

\author[0000-0002-1253-2763]{Hui Li (李辉)}
\affiliation{Department of Astronomy, Tsinghua University, Haidian DS, Beijing, 100084, People's Republic of China}

\author[0000-0003-4286-5187]{Guilin Liu (刘桂琳)}
\affiliation{CAS Key Laboratory for Research in Galaxies and Cosmology, Department of Astronomy, University of Science and Technology of China, Hefei, Anhui 230026, People's Republic of China}
\affiliation{School of Astronomy and Space Science, University of Science and Technology of China, Hefei 230026, People's Republic of China}

\author[0000-0002-4279-4182]{Paul Martini}
\affiliation{Department of Astronomy, The Ohio State University, 4055 McPherson Laboratory, 140 W 18th Avenue, Columbus, OH 43210, USA}
\affiliation{Center for Cosmology and AstroParticle Physics, The Ohio State University, 191 West Woodruff Avenue, Columbus, OH 43210, USA}
\affiliation{Department of Physics, The Ohio State University, 191 West Woodruff Avenue, Columbus, OH 43210, USA}

\author{Richard J. Rand}
\affiliation{Department of Physics and Astronomy, University of New Mexico, 210 Yale Blvd NE, Albuquerque, NM 87106, U.S.A.}

\author{Yelena Stein}
\affiliation{Ruhr University Bochum, Faculty of Physics and Astronomy, Astronimical Institute, D-44780 Bochum, Germany}

\author{Andrew W. Strong}
\affiliation{Max-Planck-Institut fuer extraterrestrische Physik, Postfach 1312, 85741 Garching, Germany}

\author[0000-0001-7936-0831]{Carlos J. Vargas}
\affiliation{Department of Astronomy and Steward Observatory, University of Arizona, Tucson, AZ, U.S.A.}

\author[0000-0002-9279-4041]{Q. Daniel Wang}
\affiliation{Department of Astronomy, University of Massachusetts, Amherst, MA 01003, U.S.A.}

\author{Jing Wang (王菁)}
\affiliation{Kavli Institute for Astronomy and Astrophysics, Peking University, Beijing 100871, People's Republic of China}

\author[0000-0002-3502-4833]{Theresa Wiegert}
\affiliation{Instituto de Astrof\'{i}sica de  Andaluc\'{i}a (IAA-CSIC), Glorieta de la Astronom\'{i}a s/n, 18008 Granada, Spain}

\author{Jianghui Xu (许蒋辉)}
\affiliation{CAS Key Laboratory for Research in Galaxies and Cosmology, Department of Astronomy, University of Science and Technology of China, Hefei, Anhui 230026, People's Republic of China}
\affiliation{School of Astronomy and Space Science, University of Science and Technology of China, Hefei 230026, People's Republic of China}

\author[0000-0001-7254-219X]{Yang Yang (杨阳)}
\affiliation{Purple Mountain Observatory, Chinese Academy of Sciences, 10 Yuanhua Road, Nanjing 210023, People's Republic of China}

\correspondingauthor{Jiang-Tao Li}
\email{pandataotao@gmail.com}

\begin{abstract}
The extraplanar diffuse ionized gas (eDIG) represents ionized gases traced by optical/UV lines beyond the stellar extent of galaxies. We herein introduce a novel multi-slit narrow-band spectroscopy method to conduct spatially resolved spectroscopy of the eDIG around a sample of nearby edge-on disk galaxies (eDIG-CHANGES). In this paper, we introduce the project design and major scientific goals, as well as a pilot study of NGC~3556 (M108). The eDIG is detected to a vertical extent of a few kpc above the disk, comparable to the X-ray and radio images. We do not see significant vertical variation of the [\ion{N}{2}]/H$\alpha$ line ratio. A rough examination of the pressure balance between different circum-galactic medium (CGM) phases indicates the magnetic field is in a rough pressure balance with the X-ray emitting hot gas, and may play an important role in the global motion of both the eDIG and the hot gas in the lower halo. At the location of an HST/COS observed UV bright background AGN $\sim29\rm~kpc$ from the center of NGC~3556, the magnetic pressure is much lower than that of the hot gas and the ionized gas traced by UV absorption lines, although the extrapolation of the pressure profiles may cause some biases in this comparison. By comparing the position-velocity diagrams of the optical and CO lines, we also find the dynamics of the two gas phases are consistent with each other, with no evidence of a global inflow/outflow and a maximum rotation velocity of $\sim150\rm~km~s^{-1}$.
\end{abstract}

\keywords{galaxies: ISM}

\section{Introduction} \label{sec:Intro}

A key component of galactic ecosystems is the circum-galactic medium (CGM), which typically distributes beyond the stellar disk and bulge of the galaxy, while still within its dark matter halo. The CGM is comprised of multi-phase gases, dust (e.g., \citealt{Whaley09}), cosmic ray (CR), and magnetic field (e.g., \citealt{Irwin12a,Irwin12b}). The multi-phase gases in the CGM include the hot gas ($T\gtrsim10^6\rm~K$) emitting X-rays (e.g., \citealt{Li13a,Li13b}), the transition-temperature gas ($T\sim10^{4-6}\rm~K$; often named ``warm-hot gas'') most commonly traced by the rest-frame UV absorption lines from high ions in the spectra of background AGN (e.g., \citealt{Chen09,Tumlinson11,Stocke13}), the $T\sim10^{3-4}\rm~K$ cool or warm gas (named differently in different studies, hereafter ``warm gas'' throughout this paper) seen in optical/UV emission lines (e.g., \citealt{Collins00,Rossa03,Haffner09,Vargas19}) or absorption lines from background AGN (e.g., \citealt{Wakker09,Werk14}), cold atomic gas often directly traced by the \ion{H}{1} 21-cm line (e.g., \citealt{Walter08,Heald11,Wang23}), and molecular gas traced by many  molecular lines typically in mm-wave (cold molecular gas, e.g., \citealt{Young95,Leroy09}) or IR (warm molecular gas, e.g., \citealt{Veilleux09}). This multi-phase gaseous CGM serves as a reservoir from which the galaxy acquires baryons to continue star formation (SF; see the recent review \citealt{FaucherGiguere23}). It also stores the kinetic energy and chemically enriched materials deposited by various types of galactic feedback, including AGN, massive stellar wind and core collapsed supernovae (SNe) from the young stellar population, or Type~Ia SNe from the old stellar population (e.g., \citealt{Strickland04a,Li13b,Li15,Wang16}). However, the astrophysics of the CGM, including the interplay among its various constituents, remains largely uncertain. 

Here we will focus on the extraplanar diffuse ionized gas (eDIG). As a key component of the multi-phase CGM, the eDIG is the $T\sim10^{3-4}\rm~K$ warm ionized gas traced by optical emission lines (e.g., H$\alpha$), and is often highly structured and distributed in a vertically extended layer above the galactic disk. The eDIG has been detected in the Milky Way via UV/optical emission (e.g., \citealt{Haffner03}) or absorption lines (e.g., \citealt{Qu18}), or pulsar dispersion measures (e.g., \citealt{Keating20}). It is also observed in highly inclined nearby disk galaxies, often via optical emission lines (e.g., \citealt{Rossa03,Vargas19}). The eDIG is a major tracer of the interface between the relatively cold galactic disk and the hotter gaseous halo. It plays a key role in the recycling of gas between the galactic disk and halo, which further affects the redistribution of gas and SF in the disk and possibly also produces extraplanar SF and high velocity clouds (e.g., \citealt{Tullmann03,Stein17}). We need to quantify the spatial distribution, physical properties, and kinematics of the eDIG in order to understand its role in the co-evolution of galaxies and the multi-phase CGM. 

The eDIG could be produced in various ways: the extended cold gas envelope photo-ionized by the ionizing radiation from the disk SF regions, the cool/warm gas outflow or the pre-existing cool CGM shock/photo-ionized by stellar and AGN feedback, the radiative cooling of the extended hot CGM, or the large scale cool gas structures photo-ionized by the UV background, etc. (e.g., \citealt{Strickland02}). Consistent with these various origins, the detected eDIG via optical emission lines can be decomposed into a few components with clearly different morphologies and extensions: (1) a thick disk or envelope with a vertical extent typically no more than a few kpc (e.g., \citealt{Lu23}), (2) some fine structures such as superbubbles and filaments directly connected to the disk SF regions with a vertical extent comparable to the thick disk (e.g., \citealt{Li09,Li19}), and (3) sometimes filaments produced by starburst driven superwinds or large scale tidal tails which could extend to $>10\rm~kpc$ (e.g., \citealt{Li08,HodgesKluck20}). There are only a few cases in which a diffuse eDIG component extending to tens of kpc has been detected around local galaxies (e.g., \citealt{Nielsen23}), but such an extended optical emission line nebula extending to even hundreds of kpc is quite common at high redshift when the Ly$\alpha$ line becomes observable from the ground (e.g., \citealt{Cantalupo14,Martin15}).

There are a few observational methods to study the eDIG around local galaxies. The most common way is the direct mapping of some optical emission lines (most commonly H$\alpha$, e.g., \citealt{Rossa03,Vargas19}), which however does not provide any direct physical and kinematic information of the gas. Long-slit spectrum with a broad wavelength coverage provides the best estimate of the physical and kinematic properties of the warm gas, but the limited spatial coverage makes it difficult to study the spatial distribution of the eDIG (e.g., \citealt{Rand97,Rand00,Boettcher19}). Imaging spectroscopy observations with Integral Field Units (IFU; e.g., \citealt{Boettcher16,Bizyaev17}) or the Fabry-P$\rm\acute{e}$rot interferometer (e.g., \citealt{Heald06a,Kamphuis07,Voigtlander13}) could simultaneously provide the spatial and spectral properties of the gas, but the disadvantage is either a small FOV and/or a narrow wavelength coverage (e.g., \citealt{Boettcher16} used an IFU with $\lesssim1\rm~arcmin^2$ FOV, $\sim5^{\prime\prime}$ angular resolution, $\sim400\rm~\AA$ wavelength coverage, and a spectral resolution of $R\sim8,000$). We therefore need a combination of some of these methods for a complete view of the eDIG.

The eDIG could also be studied via some UV lines produced by low ionization ions. Commonly used observing methods in UV band include direct imaging of UV emission lines based on special assumptions and/or instrument setups (e.g., using a combination of broad- or medium-band filters; \citealt{Hayes16,HodgesKluck20}), UV emission line spectroscopy of a small area (single aperture, no IFU, e.g., \citealt{Otte03}), UV absorption line study of a sample of background AGN (e.g., \citealt{Werk14}), and optical IFU observations of the emission line nebulae at high redshifts (e.g., \citealt{Cantalupo14,Martin15}). Many of these methods, however, are very limited in the study of local galaxies due to the lack of powerful UV telescopes. On the other hand, the study of high redshift UV emission line nebulae are often biased toward bright nebulae illuminated by an adjacent quasar, so are quite different from the eDIG discussed here.

We herein introduce the eDIG-CHANGES project, which is aimed to obtain large-FOV spatially resolved narrow-band spectroscopy of a sample of nearby edge-on spiral galaxies. This project becomes possible because of an innovative design of multi-slit masks. In \S\ref{sec:Obs}, we will introduce the project, including the sample, the design of the masks and the instrument setup, as well as our major scientific goals. We will then present a case study of NGC~3556 in \S\ref{sec:OtherDataNGC3556}, including both our multi-slit narrow-band spectroscopy observations and available multi-wavelength observations. We will further discuss the results on this case study in \S\ref{sec:ResultsDiscussions}. We will summarize our initial results and conclusions in \S\ref{sec:Summary}. Errors are quoted at $1\rm~\sigma$ confidence level throughout this paper unless specially noted.

\section{The eDIG-CHANGES project} \label{sec:Obs}

The eDIG-CHANGES project is a companion project of the Continuum Halos in Nearby Galaxies: An EVLA Survey (CHANG-ES) project \citep{Irwin12a,Irwin12b}, which mainly studies the CRs and magnetic field via radio continuum emission around 35 nearby edge-on spiral galaxies (e.g., \citealt{Wiegert15,Irwin17,Krause18,Krause20,Stein19,Stein20,MoraPartiarroyo19a,MoraPartiarroyo19b,Heald22}). The CHANG-ES sample is selected from the Nearby Galaxies Catalog (NBGC; \citealt{Tully88}), with the following criteria \citep{Irwin12a}: 1. edge-on galaxy with the inclination angle $i>75^\circ$; 2. northern sky object visible from the VLA with a declination of $\delta>-23^\circ$; 3. blue isophotal diameters $4^\prime < d_{25} < 15^\prime$; 4. 1.4~GHz radio flux density $S_{1.4}\geq23\rm~mJy$.

In the eDIG-CHANGES project, we focus on the eDIG around the same sample of galaxies as CHANG-ES. In Paper~I of the eDIG-CHANGES project \citep{Lu23} and an earlier paper \citep{Vargas19}, we present the APO 3.5m H$\alpha$ images of some CHANG-ES galaxies, and studied the vertical extent of their eDIG. But these imaging observations do not provide any spectroscopic information. In the present paper, we will introduce a new systematic spatially-resolved spectroscopic study of the eDIG around the CHANG-ES galaxies, based on a combination of our specially designed multi-slit masks, some narrow-band filters, and the OSMOS (Ohio State Multi-Object Spectrograph) imager/spectragraph mounted on the 2.4m Hiltner telescope at the MDM observatory \citep{Martini11}.

\subsection{Project design} \label{subsec:ObseDIGCHANGES}

Considering the limitations in existing eDIG observation methods stated in \S\ref{sec:Intro} (narrow-band imaging, Fabry-P\'{e}rot interferometry, long-slit or IFU spectroscopy), we design a novel method, optimized in a large FOV and narrow wavelength range. In optical band, most of the important emission lines tracing the eDIG are distributed in a few relatively narrow bands, such as the H$\alpha$ band typically covering the H$\alpha$ $\lambda 6563$~\AA\ and [\ion{N}{2}] $\lambda \lambda 6548,6583$~\AA\ lines. Therefore, we could significantly increase the efficiency of spatially resolved spectroscopy by limiting the observations only around these lines. This could be done by combining the narrow-band filter with some specially designed slit masks which are used to sample the spatial distribution of an extended source over a large FOV (e.g., \citealt{Martin08,Wu12}).

\begin{figure}[!th]
\begin{center}
\includegraphics[width=0.5\textwidth]{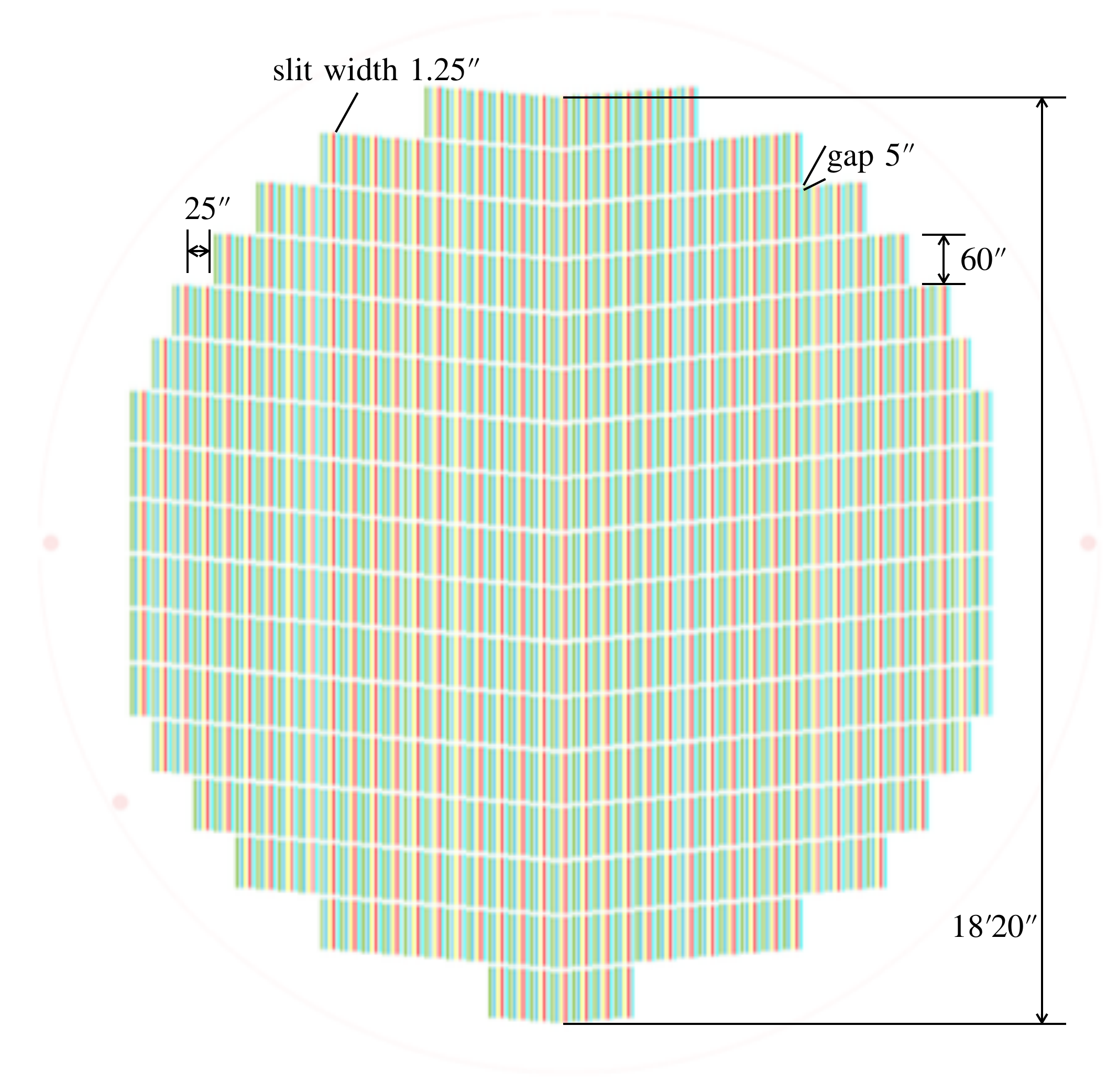}
\vspace{-0.35in}
\caption{Design of the multi-slit masks. The five different colors (green, blue, yellow, red, cyan) denote the location of the slits of the five masks. Each mask has 41 slits uniformly distributed along the horizontal direction in the figure, with a separation of $25^{\prime\prime}$. The slits run from top to bottom. Each slit has a width of $1.25^{\prime\prime}$, and is cut into multiple $60^{\prime\prime}$-long segments each gaped by $5^{\prime\prime}$ for the structural stability. The total sky coverage of the five masks with one fixed pointing is $\sim23\%$. See \S\ref{subsec:ObseDIGCHANGES} for details.} \label{fig:MultiSlitMasks}
\end{center}
\end{figure}

We design our observations based on the OSMOS imager/spectragraph mounted on the MDM 2.4m telescope \citep{Martini11}. When combined with the MDM4K CCD ($4064\times4064$), OSMOS gives a FOV of $18.5^\prime\times18.5^{\prime}$ and a pixel scale of $0.273^{\prime \prime}/$pixel at the f/7.5 focus. We design five focal plane multi-slit masks with long parallel slits to sample the spatial distribution of spectral features over almost the entire FOV. Eash mask has $41\times 1.25^{\prime \prime}$ slits uniformly spread over the focal plane, with a separation of $25^{\prime \prime}$ in the direction perpendicular to the slit to avoid overlap of the adjacent spectra (Fig.~\ref{fig:MultiSlitMasks}). The longest slit has a full length of $\approx 18.3^{\prime}$, which spread over almost the entire FOV, and is large enough for sky background subtraction for all of our sample galaxies. The slits are cut into multiple $60^{\prime \prime}$-long segments each gaped by $5^{\prime \prime}$ for the structural stability of the mask. The adjacent slits are offset by $2^{\prime \prime}$ in the direction along the slits for both the stability of the mask and the convenience of identifying the location of the spectra. 

The five multi-slit masks can be mounted in five of the six slots (one needs to leave open) in the focal plane slit wheel of OSMOS, while only one is used in a single exposure. Considering the gap between different segments, the total sky coverage of a single slit mask is $\approx 4.6\%$. Each of the five multi-slit masks are uniformly offset in the direction perpendicular to the slits by $5^{\prime \prime}$. Therefore, the total sky coverage of the five masks is $\approx 23.1\%$ in a single pointing of the telescope. We can also use the masks to conduct imaging spectroscopy observations like an IFU with almost a full sky coverage. In this case, we need four uniformly separated pointings (each shifted by one slit width of $1.25^{\prime \prime}$ in the directly perpendicular to the disk) to achieve almost a full sky coverage of 92.3\%. This mode, however, is not used quite often, because it is often difficult to point the telescope at sub-arcsec accuracy and also because the use of one or a few masks at a single pointing is often sufficient for most of our scientific goals (\S\ref{subsec:SciGoals}). 

The design of the mask is for the use together with any narrow-band filters with a typical width of $\sim 200$~\AA\ or smaller, not limited to the H$\alpha$ filter. Some H$\beta$ and [\ion{O}{3}] filters are also tested but not yet extensively used in the existing observations. The position of the spectra on the detector and the number of slits efficient in use will be different when combined with different filters and dispersers. In the present paper and a few follow-up papers, we will present results based only on our initial pilot observations in the H$\alpha$ $\lambda6563$~\AA\ band (typically also cover the two [\ion{N}{2}] $\lambda \lambda 6548,6583$~\AA\ lines). We tested a few H$\alpha$ filters available either at the MDM observatory or borrowed from the nearby Kitt Peak National Observatory (KPNO). The $4^{\prime \prime}$ KP1468 filter gives the best result: smooth line shape, ideal bandwidth of $\sim 100$~\AA\ (center wavelength $\lambda_{\rm c}=6556.02$~\AA , $FWHM=85.86$~\AA ) for a transmission $\gtrsim 30\%$ (or $\sim 200$~\AA\ for a transmission $\gtrsim 5\%$; peak transmission $T_{\rm max}\approx 70\%$; the transmission curve is available here: \url{http://mdm.kpno.noirlab.edu/filters-kpno4.html}). This filter will be used in all of our H$\alpha$ band observations.

We will use a medium resolution disperser in our multi-slit narrow-band spectroscopy observations, in order to reach a balance between the spectral resolution and wavelength coverage. The MDM observatory has two such dispersers capable with OSMOS: the VPH blue or red grism, which have a comparable spectral resolution but are sensitive in different bands. We choose the red grism in our H$\alpha$ band observations presented in this paper. The $R\approx 1600$ spectral resolution corresponds to a velocity resolution of $\Delta v\approx187\rm~km~s^{-1}$ or a wavelength resolution of $\Delta \lambda \approx 4.1$~\AA . This is not only sufficient to separate the H$\alpha$ and [\ion{N}{2}] $\lambda \lambda 6548,6583$~\AA\ lines, but also could typically result in a centroid velocity measurement accuracy of $\lesssim 50~{\rm km~s^{-1}}$ with good ${\rm S/N}$. Such a velocity resolution is sufficient to characterize the rotation curve of the galaxy at different vertical distances, especially for our edge-on $L^{\star}$ galaxy sample typically with a maximum rotation velocity of $\sim 200~{\rm km~s^{-1}}$ \citep{Irwin12a,Li16,Zheng22a}.

\subsection{Scientific goals} \label{subsec:SciGoals}

Our multi-slit narrow-band spectroscopy observations provide a unique probe of the spatial variation of optical emission line properties of the eDIG. Combined with the multi-wavelength data collected by the CHANG-ES consortium, the eDIG-CHANGES program has a few specific scientific goals which will be described in the next few subsections. Quantitative justifications of these scientific goals are largely dependent on the improvement of data calibration methods of the multi-slit narrow-band spectroscopy data, as well as the collection of multi-wavelength data and specific case studies, which will be continuously examined and published in a series of follow-up papers.

\subsubsection{Spatial variation of the physical properties of the eDIG} \label{subsubsec:SciPhysicalPropertieseDIG} 

The ultimate goal of eDIG-CHANGES is investigating warm ionized gas and resolving different processes traced by various optical lines, including H$\alpha$, H$\beta$, [\ion{O}{3}], [\ion{S}{2}], etc. The MDM observatory and KPNO have many narrow-band filters covering typical nebular emission lines, which could be used in combination with our multislit masks. We have already tested a few H$\alpha$ and [\ion{O}{3}] $\lambda 5007$~\AA\ band filters in our pilot observations. The scientific goals discussed in this subsection are largely based on multiple emission lines observed in different settings, although in this and the following several papers we will only present results based on the spectra in the H$\alpha$ band. 

The physical properties of the warm ionized gas could be measured from the ratios of some characteristic emission lines (e.g., temperature, ionization state, density, and abundance; \citealt{Osterbrock06}). For example, the [\ion{N}{2}] ($\lambda$6548+$\lambda$6583)/$\lambda$5755 and [\ion{N}{2}] $\lambda$6583/H$\alpha$ ratios are sensitive to the gas temperature (e.g., \citealt{Rand97,Boettcher19}), while in the case of the densest core of the cloudlets, the [\ion{N}{2}] $\lambda$6548/$\lambda$6583 and [\ion{S}{2}] $\lambda$6717/$\lambda$6731 ratios may be used to constrain the electron density (e.g., \citealt{Lehnert96}). Measuring the electron density may further help to constrain the magnetic field strength in combination with the rotation measure synthesis analysis from the CHANG-ES project (e.g., \citealt{Stein19,Stein20,MoraPartiarroyo19a}). Furthermore, we could also characterize the ionization conditions based on the ratios between some emission lines tracing the strength of shock ionization (e.g., [\ion{O}{3}], [\ion{N}{2}], [\ion{S}{2}]) and the Balmer lines typically tracing photo-ionization (e.g., \citealt{Lehnert96,Rand97,Rand00,Otte02}).

In most of the observations, we typically place the slit mask in a direction in which the slits are aligned perpendicular to the galactic disk. The number of slits covering the galactic disk depends on the angular size of the galaxy. For typical CHANG-ES galaxies, we could have $>10$ slits covering the galactic disk in the radial direction, and in the vertical direction the slits should be far larger than the typical extent of the eDIG. This ensures a well spatial sampling of the eDIG, even with just one slit mask (in many cases, we use more than one for a better spatial sampling). We will study the radial and vertical variations of various emission line ratios (depending on which lines are observed in a certain galaxy, at least [\ion{N}{2}] $\lambda\lambda6548,6583$ and H$\alpha$). This could help us to understand the spatial distribution of the gas and ionizing sources (e.g., \citealt{Lu23}). The measured line ratios at different vertical and radial distances will also be systematically compared to the SFR or the SFR surface density of the galaxies \citep{Vargas19}, in order to examine the relation between the eDIG and the SF feedback (e.g., \citealt{Lehnert96}).

\subsubsection{Kinematics of the eDIG} \label{subsubsec:SciKinematicseDIG} 

The multi-slit narrow-band spectroscopy mode is very efficient in constructing rotation curves at different vertical distances from the galactic plane in edge-on disk galaxies. The spectral resolution of $\Delta v\approx187\rm~km~s^{-1}$ is typically insufficient to measure the width of the emission lines from the eDIG, but with good S/N it is often sufficient to measure the centroid velocity of the line to an accuracy of $<50\rm~km~s^{-1}$ (see results in \S\ref{subsec:ResultseDIG}). This is typically good enough to characterize the rotation curve in an $L^\star$ galaxy with a maximum rotation velocity of $\sim200\rm~km~s^{-1}$. 

The ``lagging'' gaseous halo in comparison to the underlying galactic disk is a powerful probe of the gas origin and dynamics around a disk galaxy. The lagging halo is most commonly observed in cold gas via the \ion{H}{1} 21cm line (e.g., \citealt{Swaters97,Schaap00,Fraternali06,Oosterloo07}), and less well studied in higher temperature ionized gas because of the often poorer spatial or spectral sampling (with long-slit spectroscopy or IFU which have limited FOV, or Fabry-P$\rm\acute{e}$rot interferometer with limited wavelength coverage, e.g., \citealt{Tullmann00,Fraternali04,Heald06a,Heald07,Kamphuis07,Voigtlander13}). The lagging halo is often interpreted as the cold gas envelope around disk galaxies being affected by the galactic fountain or the externally accreted gas, which produces the radial variation of the vertical gradient of the rotation velocity (e.g., \citealt{Oosterloo07,Zschaechner15}). Alternative interpretations of the rotational velocity lag may involve the large-scale magnetized gas (e.g., \citealt{Benjamin02,Henriksen16}). The ionized gas, on the other hand, may have significantly different origins and dynamics than the cold gas, such as being more severely affected by the hot galactic wind (e.g., \citealt{Boettcher19}). We need a better spatial sampling of the eDIG in spectroscopy observations, in order to better understand both the lagging halo seen in atomic and ionized gas and the possibly more complicated dynamics of the higher temperature ionized gas.

We will characterize the lagging gaseous halo by constructing and analyzing the rotation curves of the optical emission lines at different vertical heights from the galactic plane. Such studies have only been conducted in a few galaxies, mostly based on the H$\alpha$ line tracing the relatively cool photo-ionized gas (e.g., \citealt{Heald06a,Kamphuis07}). We will measure the vertical gradient of the rotation velocity and systematically compare it to those predicted by different models, such as a pure ballistic (e.g., \citealt{Collins02}) or hydrostatic model (e.g., \citealt{Barnabe06}). Such comparisons are often efficient ways to search for signatures of unusual gas dynamical structures, such as gas accretion from the CGM (e.g., \citealt{Fraternali06}). In addition to the 1D rotation curves, we will also construct position-velocity (PV) diagrams for different emission lines and at different distances from the disk. We will further compare the PV diagram on the galactic plane to those measured in other gas probes, such as our IRAM 30m CO observations (with a few pointings spread over the galactic disk; e.g., \citealt{Jiang23}), in order to search for features of coupled warm/cold gas flows at the disk/halo interface. 

\subsubsection{Pressure balance of different CGM phases} \label{subsubsec:SciPressureBalanceCGM}

We will further compare the eDIG to other CGM phases based on the multi-wavelength data collected by the CHANG-ES consortium (e.g., \citealt{Irwin12a,Irwin12b,Li13a,Li13b,Vargas19,Jiang23}), in order to examine the pressure balance between different CGM phases. Such a comparison is the key to understand two important scientific problems: (1) What is the supporting mechanism of the CGM? (2) What is the driving force of the galactic outflow? 

There are various types of gas flows at the disk halo interface: hydrostatic equilibrium, galactic fountain, galactic wind, or accretion flow (e.g., \citealt{Bregman80}). The key parameter determining the dynamical state of the extraplanar gas is the relative importance of different supporting pressures, including but not limited to the thermal, turbulent, radiation, magnetic, and CR pressures. Comparing the vertical extents of the eDIG to other CGM phases is in general an efficient way to justify the dominance of different pressures at different vertical distances (e.g., \citealt{Rand97,Lu23}). 

The supporting pressure of the eDIG could be further studied by comparing the vertical pressure gradient of different CGM phases to the gravitational potential of the galaxy (e.g., \citealt{Boettcher16,Schmidt19}). The pressure of different CGM phases could be measured with multi-wavelength observations. For example, in optical emission line observations, when the temperature of the warm ionized gas is poorly constrained (e.g., when the spectral resolution is too poor to measure the line profile), the thermal pressure of the eDIG is often calculated by adopting the measured gas density profile and an assumed warm gas temperature of $T\sim10^4\rm~K$ (e.g., \citealt{Boettcher16}). The turbulent pressure could be estimated from the width of the emission lines in high-resolution spectra, after subtracting the velocity dispersion expected from the instrument broadening, the thermal motions, and the global galactic rotation (e.g., \citealt{Boettcher16}). Furthermore, optical emission line observations could also be used to constrain the radiation pressure mainly contributed by young stars from the galactic disk (e.g., \citealt{Lopez14,DellaBruna22}). On the other hand, radio continuum observations (mainly synchrotron emission from CR electrons in the galactic magnetic field) could be used to measure the magnetic and CR pressures in the galactic halo \citep{Irwin12a,Irwin12b,Schmidt19}, while X-ray observations (mainly line emissions from hot plasma in thermal equilibrium) could help us to measure the thermal pressure of the hot gas (e.g., \citealt{Wang01,Li13b}). 

There are many galaxies in the CHANG-ES sample showing signatures of outflows on different physical scales, such as jet, superbubble, and large-scale ``X''-shaped superwind structures (e.g., \citealt{Li08,Li19,Li22,Krause18,Krause20,Miskolczi19,Stein20,HodgesKluck20,Heald22,Stein23}). Another key science related to the pressure balance of different CGM phases is to understand the mechanisms driving these outflows. Galactic outflows can be driven in different ways, such as by thermal pressure (e.g., \citealt{Chevalier85}), radiation pressure (e.g., \citealt{Krumholz12}), momentum (e.g., \citealt{Murray05}), or CRs and MHD waves (e.g., \citealt{Breitschwerdt91,Henriksen21}). 

Based on the multi-wavelength data collected by the CHANG-ES consortium, there are a few ways to understand which of these mechanisms play a key role in driving the outflows: (1) We can directly compare the pressure of different CGM phases entrained in the outflow (or enclosed by the outflow structures; e.g., \citealt{Boettcher16,Boettcher19,Li22}). These pressures include the thermal and turbulent pressures of the warm ionized gas, the thermal pressure of the hot gas, the CR and magnetic pressures, and the radiation pressure calculated indirectly from the SFR. (2) We will also quantify the galactic feedback energy budget by comparing the energy injected by different types of feedback (AGN, young and old stellar populations) to the energy stored in different CGM phases (e.g., \citealt{Li13b,Li16}). There is always a small fraction of feedback energy detected in a single CGM phase, resulting in the ``missing galactic feedback problem'' (e.g., \citealt{Li20}). A complete survey of the feedback energy budget in different CGM phases is thus critical in understanding the feedback processes. (3) In some special cases where a galaxy hosts a well defined superbubble (e.g., \citealt{Li19,Li22}), we can also compare the observed size and expansion velocity of the bubble to analytical models, in order to estimate the required energy injection rate to produce the bubble (e.g., \citealt{Weaver77,MacLow88}). This method helps to justify if starburst or AGN feedback can be responsible for the blowout of the bubble.

\section{A Pilot Case Study of the Multi-phase CGM of NGC~3556} \label{sec:OtherDataNGC3556}

We choose the galaxy NGC~3556 (M108) for a pilot case study of the multi-phase CGM. This multi-wavelength study invokes our new multi-slit narrow-band spectroscopy data from the eDIG-CHANGES project. NGC~3556 is a nearby edge-on (inclination angle $i=81^\circ$) field (density of galaxies brighter than -16~mag in the vicinity $\rho=0.15$) SBc galaxy at a distance of $d=14.1\rm~Mpc$ ($1^{\prime\prime}\approx69\rm~pc$), showing a systemic velocity of $699\rm~km~s^{-1}$ \citep{Irwin12a}. NGC~3556 is typical among all CHANG-ES galaxies, considering its angular size ($D_{\rm 25}=7.8^\prime$), star formation properties (revised SFR using both the IR and H$\alpha$ luminosities $\rm SFR=3.57~M_\odot~yr^{-1}$; SF surface density computed with the SFR and the IR diameter $\rm SFR_{SD}=7.32\times10^{-3}~M_\odot~yr^{-1}~kpc^{-2}$), mass (rotation velocity $v_{\rm rot}=153.2\rm~km~s^{-1}$; the photometric mass calculated from the 2MASS K-band luminosity $M_*$ and the dynamical mass calculated from $v_{\rm rot}$ are both $\approx2.8\times10^{10}\rm~M_\odot$), and radio properties (6~GHz flux density $f_{\rm 6GHz}=79.2\rm~mJy$; 1.5~GHz flux density $f_{\rm 1.5GHz}=291.5\rm~mJy$; \citealt{Irwin12a,Wiegert15,Li16,Vargas19}). As a pilot study, these moderate properties reduce the biases in any extreme conditions.

\begin{figure*}[!th]
\begin{center}
\includegraphics[width=1.0\textwidth]{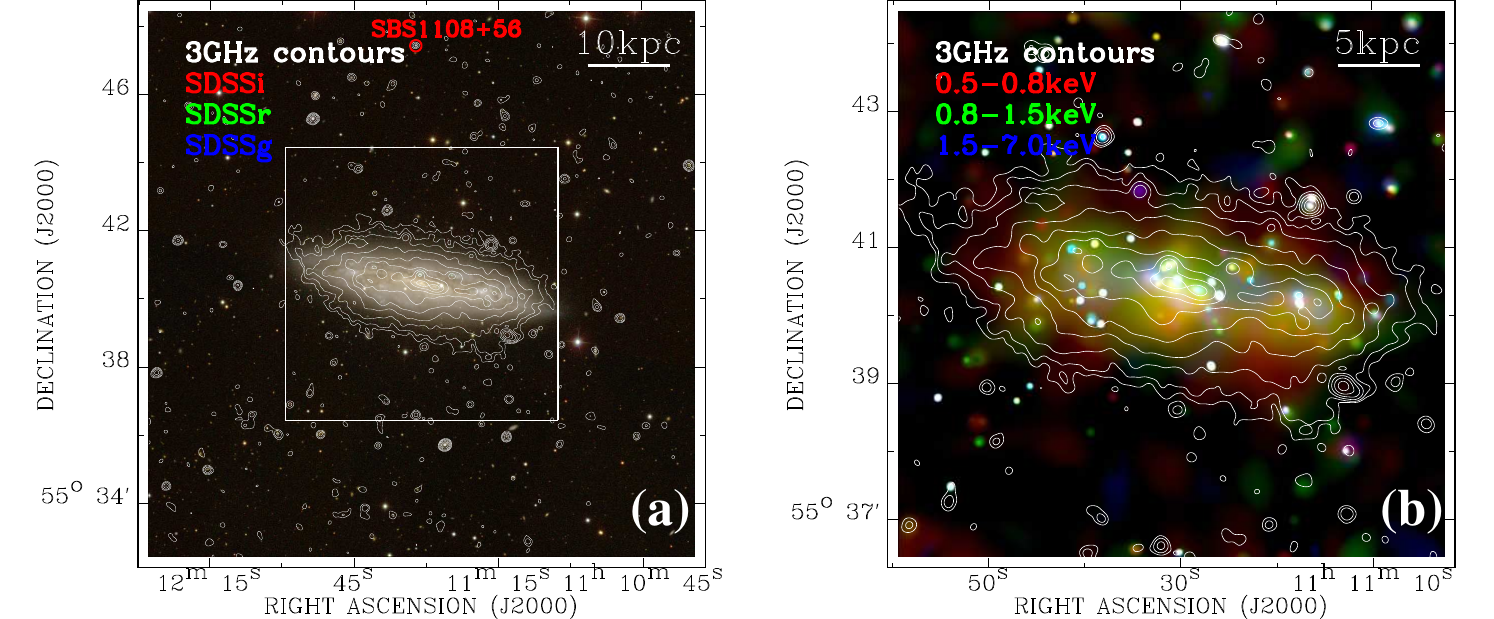}
\caption{Multi-wavelength images of NGC~3556. (a) $16^\prime\times16^\prime$ SDSS i-, r-, g-band tri-color images. The white contours are the $10^{\prime\prime}$ uv-tapered \emph{VLA} C-configuration S-band (3~GHz) image \citep{Xu24}. The radio continuum emission has been detected above the optical disk of the galaxy. The small red circle is the UV-bright background AGN SBS 1108+560 observed by the \emph{HST}/COS (\S\ref{subsec:OtherDataHSTCOS}). This AGN is also detected in radio. The white box is the FOV of the \emph{Chandra} images shown in panel~(b). (b) $8^\prime\times8^\prime$ background-subtracted, vignetting-corrected, adaptively smoothed \emph{Chandra} tri-color images (0.5-0.8, 0.8-1.5, 1.5-7~keV; \citealt{Li13a}) overlaid with the same radio contours as in panel~(a). The X-ray and radio emissions have similar morphologies.
}\label{fig:MultiBandImgNGC3556}
\end{center}
\end{figure*}

\begin{figure*}[!th]
\begin{center}
\epsfig{figure=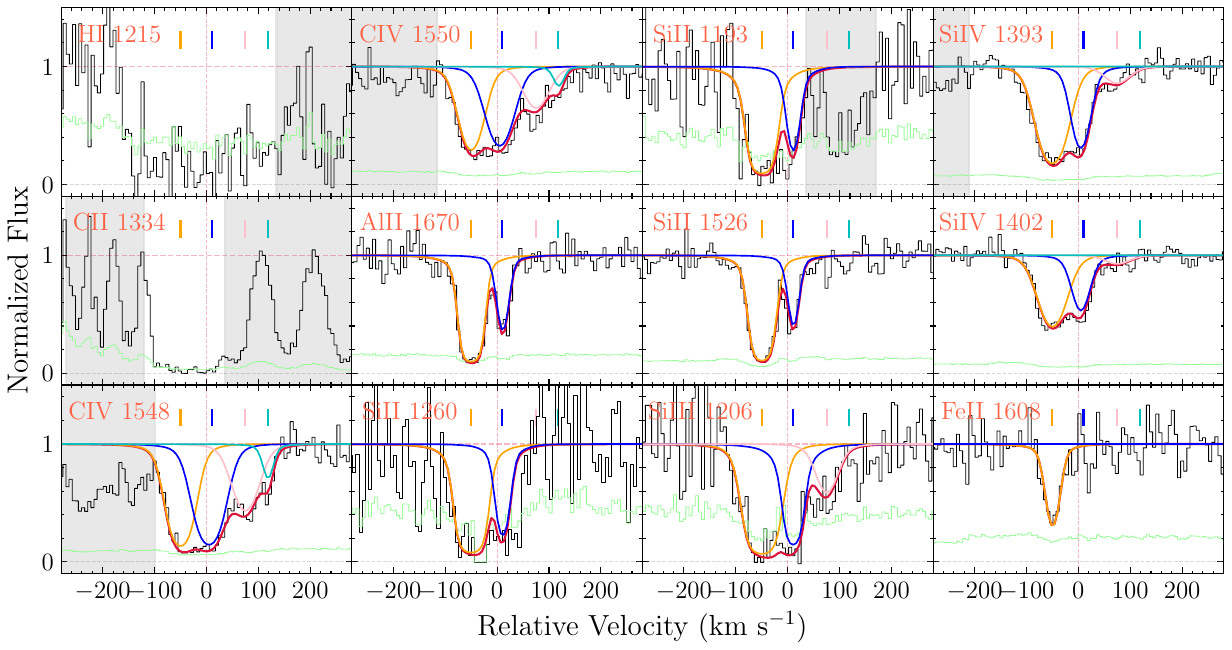,width=0.95\textwidth,angle=0}
\caption{Identified absorption lines in the \emph{HST}/COS spectra of SBS~1108+560, at a receding velocity close to $V_{\odot}\approx 699~{\rm km~s^{-1}}$ associated with NGC~3556. The black curve in each panel is the COS spectrum, while the red curve is the total fitted model. Absorption features are decomposed into four components with different line centroids: orange at $-50~{\rm km~s^{-1}}$; blue at $+10~{\rm km~s^{-1}}$; pink at $+75~{\rm km~s^{-1}}$; and cyan at $+118~{\rm km~s^{-1}}$, from a systemic velocity of $699~{\rm km~s^{-1}}$.}\label{fig:HSTCOSSpec}
\end{center}
\end{figure*}

As a well studied nearby galaxy, there are many multi-wavelength data available for NGC~3556 (e.g., Fig.~\ref{fig:MultiBandImgNGC3556}). Vertically extended H$\alpha$ \citep{Collins00,Vargas19}, X-ray \citep{Wang03,Li13a}, radio continuum \citep{Wiegert15,Krause20}, and \ion{H}{1} 21-cm line \citep{Zheng22a,Zheng22b} emissions have been confirmed in previous studies (see a comparison of their scale heights in \citealt{Lu23}). Furthermore, as a very rare case in the CHANG-ES sample, there is also a UV bright background AGN SBS 1108+560 projected at $\approx29\rm~kpc$ north above the galactic disk of NGC~3556 (Fig.~\ref{fig:MultiBandImgNGC3556}a). The archival HST/COS observations of this AGN could help us to study the warm ionized gas around this galaxy via UV absorption lines (Fig.~\ref{fig:HSTCOSSpec}). In the following subsections, we will briefly introduce these multi-wavelength data.

\subsection{Pilot multi-slit narrow-band spectroscopy observations of NGC~3556} \label{subsec:ObsNGC3556}

\begin{figure*}[!th]
\begin{center}
\epsfig{figure=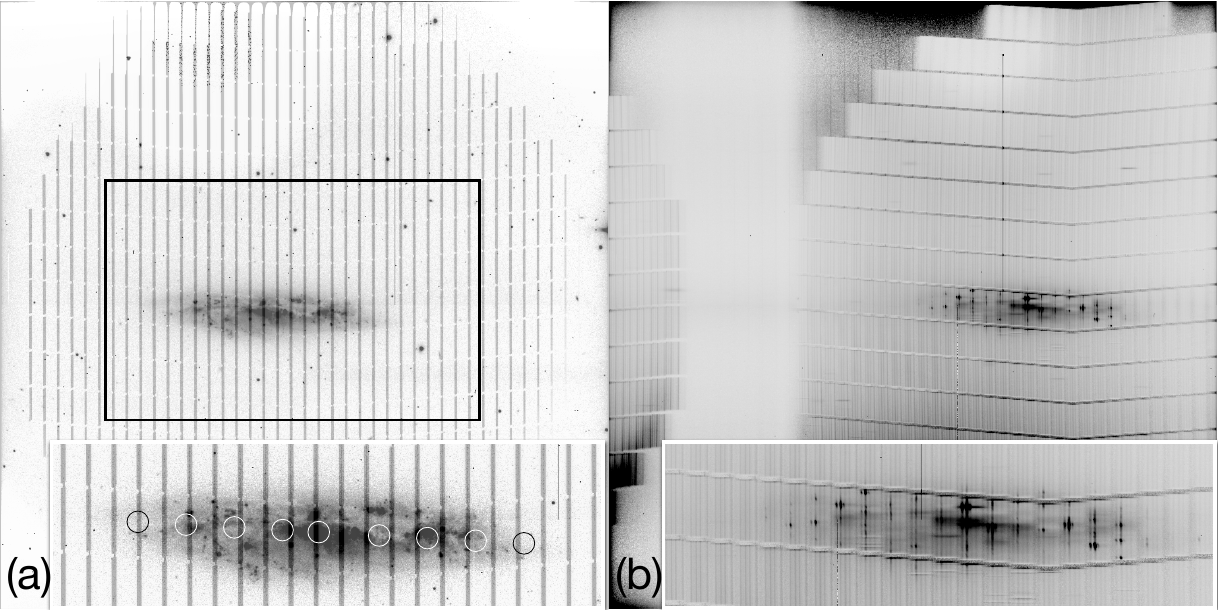,width=0.95\textwidth,angle=0, clip=}
\caption{Raw images from the multi-slit narrow-band spectroscopy of NGC~3556. The intensity scale is inverted such that bright emission is black. The instrument (OSMOS) has been rotated for $-7.5^{\circ}$ so the galactic disk is placed perpendicular to the slits. (a) Slit image overlaid on a snapshot H$\alpha$ image, taken with the same filter KP1468 as the spectroscopy observations, and with an exposure of $60$~sec. The black box indicates the region shown in Fig.~\ref{fig:MultiSlitParaMapNGC3556}. The inserted panel shows a zoom-in of the central region, in which the circles mark the locations of the IRAM 30m beams with a diameter of 21.39$^{\prime \prime}$ at the $^{12}$CO $J=1-0$ band \citep{Jiang24}. These beams are used to construct the position-velocity (PV) diagram in Fig.~\ref{fig:PVDiagram}b. (b) Raw spectra taken with the multi-slit mask and the KP1468 H$\alpha$ narrow-band filter. The image in the leftmost side should be higher order spectra of the rightmost slits, which are not used in the follow-up analysis. Each rectangle, outlined in white edges, is a narrow-band spectrum with the $60^{\prime\prime}$ slit segment, that traces spatial length, in the vertical direction and wavelength spanning the slimmer, horizontal direction. The inserted panel shows a zoom-in of the central region, where the three prominent emission lines (H$\alpha$, [\ion{N}{2}]$\lambda\lambda6548,6583$~\AA ) can be clearly seen as bright vertical lines.}\label{fig:MultiSlitRawImgNGC3556}
\end{center}
\end{figure*}

The multi-slit narrow-band spectroscopy data of NGC~3556 used in the present paper was taken in a test run at the MDM observatory with the 2.4m Hiltner telescope and the OSMOS imager/spectrograph on January 25, 2020. Similar data is continuously obtained via the eDIG-CHANGES project, so there will be more data of this galaxy published in future papers, specifically with a better spatial sampling along the galactic disk direction employing all the five slit masks. We herein focus on data taken just in this night as a simple example. The instrument was rotated by $-7.5^\circ$, so the long slits is aligned in a direction perpendicular to the disk of NGC~3556 (position angle $PA=82.5^\circ$ from north and measured counterclockwise). The observations were taken with the slit mask located at just one position. In total about 17-18 of the 40 long slits cover the optical disk of the galaxy (Fig.~\ref{fig:MultiSlitRawImgNGC3556}). This configuration provides sufficient sampling to extract the rotation curves at different vertical distances. We took $10\times1800\rm~s$ exposures (in total 5~hours) with the multi-slit mask, the KP1468 H$\alpha$ narrow-band filter, and the MDM red grism (\S\ref{subsec:ObseDIGCHANGES}). 

\begin{figure*}[!th]
\begin{center}
\includegraphics[width=1.0\textwidth]{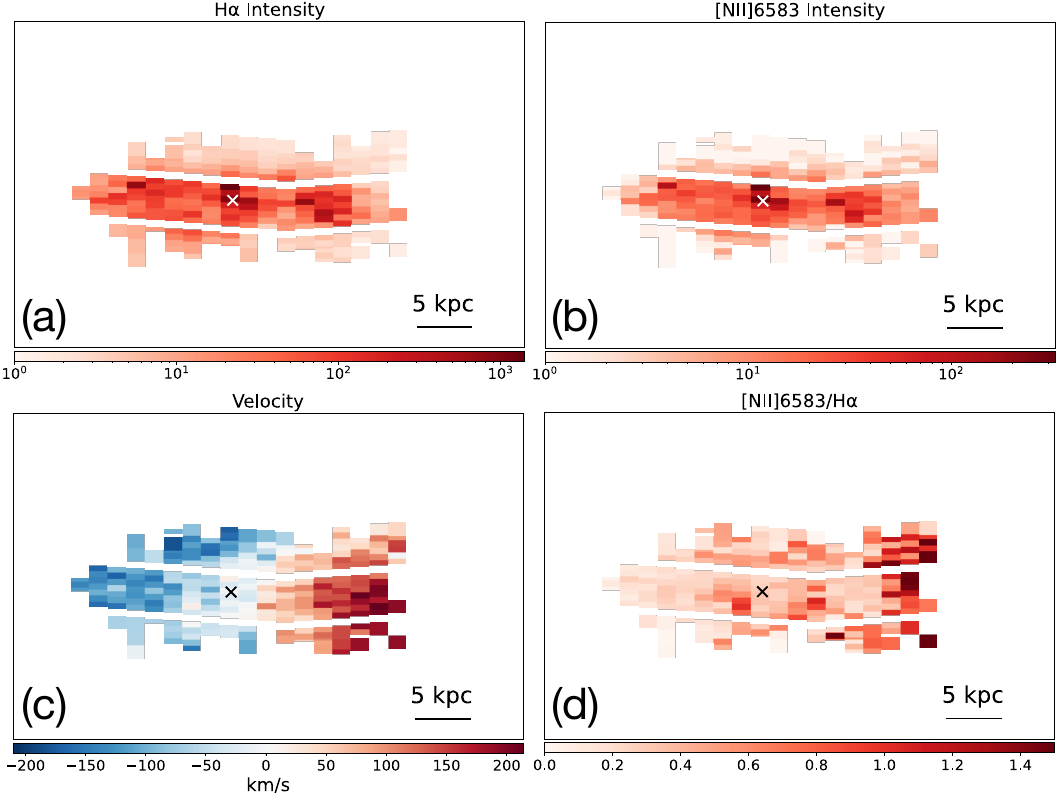}
\caption{Calibrated data and initial results from the multi-slit spectroscopy observations. (a) and (b) show the H$\alpha$ and [\ion{N}{2}]$\lambda 6583$~\AA\ line intensities in arbitrary unit. (c) shows the line centroid velocity jointly estimated from both the H$\alpha$ and the [\ion{N}{2}]$\lambda 6583$~\AA\ lines. (d) shows the [\ion{N}{2}]$\lambda 6583$~\AA -to-H$\alpha$ line intensity ratio. The white or black cross in each panel marks the galactic center.}\label{fig:MultiSlitParaMapNGC3556}
\end{center}
\end{figure*}

Procedures reducing the multi-slit narrow-band spectroscopy data will be described in detail in \citet{Lu24}. We herein just present some initial results, including the H$\alpha$ and [\ion{N}{2}] flux images (no flux calibration), the [\ion{N}{2}]/H$\alpha$ line ratio image, and the line centroid velocity image (Fig.~\ref{fig:MultiSlitParaMapNGC3556}). The shape of the galactic disk is clearly shown in the line images, and the [\ion{N}{2}]/H$\alpha$ line ratio shows no significant vertical gradient. Furthermore, the galaxy also shows a clear rotating disk on the line centroid velocity map. These initial results confirm the reliability of the proposed approach of the multi-slit narrow-band spectroscopy observations as described in the above sections (\S\ref{subsec:ObseDIGCHANGES}). However, because of the limited spectral resolution of $R\sim1600$ and the narrow wavelength range of $\sim100$~\AA , we do the wavelength calibration only based on a few lamp lines in a narrow-wavelength range (with some additional corrections using the sky lines and observations of a planetary nebula as the wavelength calibration source; eDIG-CHANGES~III). The resultant line centroid velocity map may thus be not quite accurate (the absolute velocity typically has an uncertainty of a few tenth of $\rm~km~s^{-1}$, see \citealt{Lu24} for more discussions). However, as shown in Fig.~\ref{fig:MultiSlitParaMapNGC3556}c, we see clear rotating disk features with a typical maximum rotation velocity of $v_{\rm rot}\sim150\rm~km~s^{-1}$, comparable to those determined with other methods (e.g., see the catalogue listed in \citealt{Li13a}). We therefore believe the wavelength calibration is in general reliable and the spectral resolution is sufficient to characterize the rotation curve of $L^\star$ galaxies.

\subsection{HST/COS observations of the UV-bright background AGN SBS 1108+560} \label{subsec:OtherDataHSTCOS}

SBS~1108+560 is a UV-bright background AGN with a redshift of $z\approx 0.765$, located at RA $=$ 11$^{\rm h}$11$^{\rm m}$32$^{\rm s}$.19 and DEC $=+$55$^{\rm d}$47$^{\rm m}$26$^{\rm s}$.08. Its \emph{GALEX} far-UV magnitude is $m_{\rm FUV}=17.5$, indicating a flux of $F_{\rm FUV}=4.7\times 10^{-15}~{\rm ergs~s^{-1}~cm^{-2}~\AA ^{-1}}$ at $1540$~\AA . In the CHANG-ES sample, SBS~1108+560 is one of the UV brightest AGN with the smallest projected distance (impact parameter) from the galactic nucleus ($\rho \approx 6.99^{\prime}\approx 29$~kpc from NGC~3556; Fig.~\ref{fig:MultiBandImgNGC3556}a), providing us with a unique opportunity to study the UV absorption lines arising from the galactic disk-halo interface (e.g., \citealt{Bregman13, Qu19}). SBS~1108+560 is also detected in radio (\citealt{Xu24}; Fig.~\ref{fig:MultiBandImgNGC3556}a), but not covered by the archival \emph{Chandra} observations (Fig.~\ref{fig:MultiBandImgNGC3556}b).

SBS 1108+560 was observed with the FUV gratings G130M and G160M of \emph{HST}/COS for total exposure times of $\approx8.39\rm~ks$ and $\approx8.85\rm~ks$ on May 12th, 2012 (COS-GTO; PID$=$12025, PI: Green). The medium-resolution (FWHM $\approx 20\rm~ km~s^{-1}$) {\it HST}/COS spectra were coadded using custom software to optimize the S/N ratio and wavelength calibration over a contiguous spectral coverage from $1100$~\AA\ to $1800$~\AA\ \citep{Qu19}. At $z\approx 0$, the {\it HST}/COS FUV spectra cover a wide range of species, including \ion{H}{1}, \ion{C}{2}, \ion{C}{4}, \ion{N}{5}, \ion{Al}{2}, \ion{Si}{2}-{\small IV}, and \ion{Fe}{2}.

We search absorption features of these species in a velocity window of $\pm~ 500\rm~km~s^{-1}$ around the systemic velocity of NGC~3556. For detected absorption transitions, we perform a Voigt profile analysis to decompose absorption components and determine the column density ($N$), line centroid velocity ($v_{\rm c}$), and Doppler width ($b$), following the procedure described in \citet{Qu22}. The fitting results are plotted in Fig.~\ref{fig:HSTCOSSpec} and summarized in Table~\ref{table:NionCOS}. The reported measurements are median with $1~\sigma$ uncertainties, while $2~\sigma$ lower limits are reported for saturated transitions.

Using the measured column density of different ions, we run a series of photoionization models with different hydrogen column densities $\log N_{\rm H}/{\rm cm^{-2}}$ (18.5-18.9), radiation fields (with or without galactic radiation field), and temperatures [$(0.5-1)\times10^4\rm~K$] to constrain the physical properties of the UV-absorbing gas. These models are obtained using the Cloudy (v17; \citealt{Ferland2017}). In those models, the low ions (e.g., \ion{Si}{2}, \ion{Fe}{2}, and \ion{Al}{2}) and high ions (e.g., \ion{C}{4} and \ion{Si}{4}) cannot be fitted with a single phase photoionization model, assuming the solar abundance.

\begin{figure}
\begin{center}
\epsfig{figure=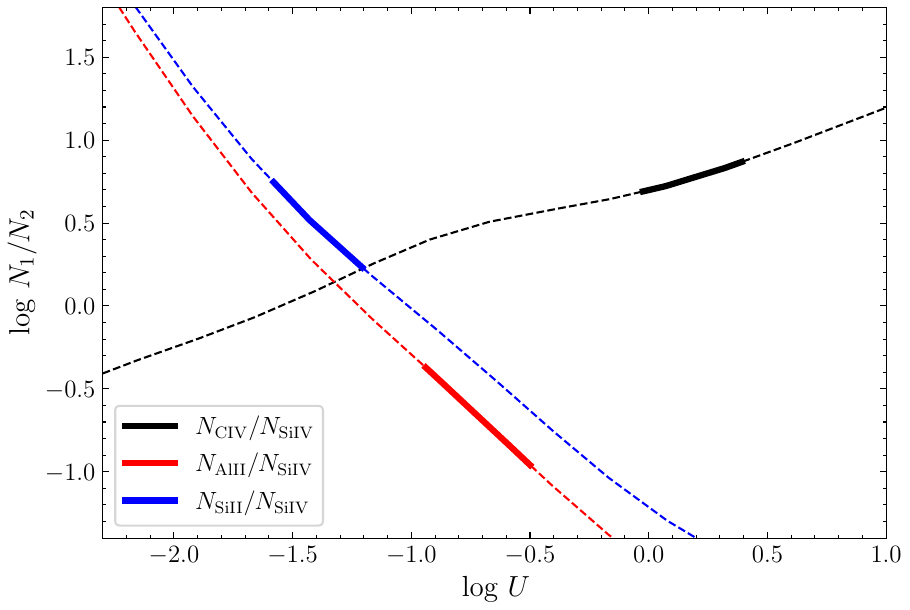,width=0.48\textwidth,angle=0}
\caption{An example of photoionization modeling of the absorption component at $\approx +10\rm~km~s^{-1}$, assuming solar metallicity. Column density ratios of different ions indicate different ionization parameters, which cannot be modeled by a single-phase photoionization model.}
\label{fig:cos_cloudy}
\end{center}
\end{figure}

We present an example of such photoionization models for the $v_{\rm c}\approx +10 \rm~km~s^{-1}$ component in Fig.~\ref{fig:cos_cloudy}. In this calculation, we adopt the UV background model from \citet{FaucherGiguere20}, and assume the solar abundance pattern and photoionization equilibrium. Using the four ions with measured column densities from unsaturated transitions (\ion{Si}{2}, \ion{Al}{2}, \ion{Si}{4}, and \ion{C}{4}), we calculate the column density ratio between different ions as a tracer of the ionization parameter $U$. These ratios cannot be produced with a single ionization parameter, suggesting the need for multiphase models under the assumption of photoionization equilibrium. Various factors may affect the ionization modeling, including additional heating/ionizing sources or the significant non-solar abundance pattern. At the current moment, a solid conclusion considering these uncertainties is limited by the quality of the available data.

\begin{table}
\begin{center}
\caption{Properties of different absorption components along the background AGN SBS~1108+560}
\footnotesize
\tabcolsep=7.0pt%
\begin{tabular}{lccc}
\hline\hline
 Ion & $\log N/{\rm cm^{-2}}$ & $b/\rm km~s^{-1}$ & $v_{\rm c}/\rm km~s^{-1}$ \\
\hline 
\ion{H}{1}$^{a}$ & $>14.8$ & $...$ & $...$ \\
\hline
\ion{C}{2}$^{a}$ & $>15.5$ & $...$ & $...$ \\
\hline
\ion{C}{4} & $14.68\pm0.03$ & $...$ & $...$ \\
\cline{2-4}
& $14.30\pm 0.10 $ & $25.4\pm 4.0 $ & $-50.3\pm 3.8 $\\
& $14.29\pm 0.09 $ & $29.3\pm 6.1 $ & $4.5\pm 3.7 $\\
& $13.85\pm 0.09 $ & $29.1\pm 7.9 $ & $74.6\pm 4.4 $\\
& $13.16_{-0.33}^{+0.17} $ & $9.2\pm 6.0 $ & $118.0\pm 5.0$\\
\hline
\ion{Si}{2} & $>14.6$ & $...$ & $...$ \\
\cline{2-4}
& $>14.5 $ & $14.1\pm 3.1 $ & $-50.0\pm 1.2 $\\
 & $13.88_{-0.14}^{+0.44} $ & $8.4\pm 3.5 $ & $10.4\pm 2.0 $\\
\hline
\ion{Si}{3} & $>13.7$ & $...$ & $...$ \\
\cline{2-4}
& $>13.2 $ & $23.9\pm 11.1 $ & $-50.0$$^{b}$ (\ion{Si}{2})\\
 & $>12.9 $ & $8.4$$^{b}$ (\ion{Si}{2}) & $10.4$$^{b}$ (\ion{Si}{2})\\
 & $>12.0 $ & $29.1$$^{b}$ (\ion{C}{4}) & $75.0$$^{b}$ (\ion{C}{4})\\
\hline
\ion{Si}{4} & $14.02\pm0.02$ & $...$ & $...$ \\
\cline{2-4}
& $13.83\pm 0.05 $ & $30.7\pm 4.6 $ & $-50.3$$^{b}$ (\ion{C}{4})\\
& $13.51\pm 0.10 $ & $18.8\pm 4.6 $ & $4.5$$^{b}$ (\ion{C}{4})\\
& $12.70\pm 0.12 $ & $29.1$$^{b}$ (\ion{C}{4}) & $74.6$$^{b}$ (\ion{C}{4})\\
\hline
\ion{Al}{2} & $>13.4$ & $...$ & $...$ \\
\cline{2-4}
& $>13.4 $ & $11.8\pm 2.9 $ & $-50.0$$^{b}$ (\ion{Si}{2})\\
 & $12.72_{-0.17}^{+0.42} $ & $10.2\pm 3.9 $ & $10.4$ $^{b}$(\ion{Si}{2})\\
 \hline
\ion{Fe}{2} & $14.35_{-0.18}^{+0.38}$ & $...$ & $...$\\
\cline{2-4}
& $14.35_{-0.18}^{+0.38}$ & $12.8\pm 6.2 $ & $-50.0$$^{b}$ (\ion{Si}{2})\\
 & $<13.9$ & $8.4$$^{b}$
(\ion{Si}{2}) & $10.4$$^{b}$
(\ion{Si}{2})\\
\hline\hline
\end{tabular}\label{table:NionCOS}
\end{center}
The first row of each ion indicate the total column density of different resolved components, which are adopted in Fig.~\ref{fig:PressureBalanceXrayHighLowIons}. \\
$^{a}$ Absorption features associated with these two ions are not decomposed because of significant contamination or low S/N. \\
$^{b}$ These parameters are tied to the relevant ions labeled in the table, leading to better decomposition.\\
\end{table}

\subsection{X-ray and radio data} \label{subsec:OtherDataCGM}

We use the X-ray and radio data to trace the hot gas and magnetic field/CRs (also see \S\ref{sec:ResultsDiscussions} for further discussions). The \emph{Chandra} X-ray observation of NGC~3556 was taken in September 2001 for a total exposure time of $\approx60\rm~ks$ (Fig.~\ref{fig:MultiBandImgNGC3556}b). The \emph{Chandra} data was first published in \citet{Wang03}. We herein directly adopt the reduced data from \citet{Li13a,Li13b,Li14}, which has an effective exposure time of $57.98\rm~ks$ after removing background flares. The data reduction procedures, including basic calibrations, detecting and removing point-like sources, as well as spectral and spatial analysis of the diffuse X-ray emission, are detailed in \citet{Li13a}. In the present paper, we will need the hot gas density and temperature from this \emph{Chandra} observation to calculate its thermal pressure, in order to examine the pressure balance between different CGM phases (\S\ref{sec:ResultsDiscussions}).

\citet{Li13a} presented analysis of the \emph{Chandra} X-ray spectrum extracted from a vertical range of $\mid h\mid \lesssim 1.55^{\prime}\sim 6.5$~kpc above the disk of NGC~3556. The average electron density of the hot gas is $n_{\rm e}\approx 2.5\times 10^{-3}f_{\rm X}^{-1/2}~{\rm cm^{-3}}$, where $f_{\rm X}$ is the volume filling factor of the X-ray emitting gas and is simply assumed to be $f_{\rm X}\sim 1$ throughout the paper. In order to compare to the UV absorption line measurements (\S\ref{subsec:OtherDataHSTCOS}), we also need to roughly estimate the hot gas properties at the location of the background AGN. SBS~1108+560 is at a vertical distance of $\sim 7^{\prime}$ above the galactic disk. Considering the non-detection of the soft X-ray excess, the possible contribution of disk emission in the \emph{Chandra} spectral analysis, and the typical slope of the X-ray intensity profile of the hot CGM (e.g., for a $\beta$-model with a slope of $\beta =0.5$; \citealt{Li17}), we simply adopt a hot gas electron density of $n_{\rm e}=3\times 10^{-4}f_{\rm X}^{-1/2}~{\rm cm^{-3}}$ at the location of this background AGN in the following discussions. Such an assumed $n_{\rm e}$ value is only valid on order of magnitude. We also assume no variation of the hot gas temperature from the spectral analysis region to the location of the background AGN, i.e., $kT\approx 0.3$~keV throughout the halo \citep{Li13a}.

The radio continuum data of NGC~3556 is obtained from the CHANG-ES program \citep{Irwin12a,Irwin12b}. The \emph{VLA} C- and L-band and the \emph{LOFAR} low frequency data of NGC~3556 from CHANG-ES were published in \citet{Wiegert15,Miskolczi19,Krause20}. The new S-band data of the CHANG-ES sample was obtained through a follow-up \emph{VLA} program (approved in 2020A, 2021A, 2022B, PI: Yelena Stein). The C-configuration S-band image of NGC~3556 is shown in Fig.~\ref{fig:MultiBandImgNGC3556}, and will be published in \citet{Xu24}. These radio continuum observations mainly trace the synchrotron emission of CR electrons produced in the galactic magnetic field. 

We herein need to estimate the magnetic and CR pressures to be compared to the thermal pressures of the multi-phase gases. In this paper, we adopt the magnetic pressure derived only from the \emph{VLA} S-band observations under the energy equipartition assumption, i.e., the energy density (so pressure) of the CR and magnetic field equal to each other \citep{Xu24}. Compared to the C- and L-band data, the S-band data is at a good balance between the Faraday depolarization and resolution, and is typically a crucial transition zone between Faraday thin and Faraday thick regimes, so optimized for rotation measure synthesis and magnetic field measurement. The energy equipartition assumption is often reliable on large enough physical scales (e.g., a few kpc), where the CR electrons and magnetic field could reach an equilibrium state (e.g., \citealt{Beck05}). We simply adopt a magnetic field strength of $B\approx 5\rm~\mu G$, which is an average value calculated in the X-ray spectral analysis region \citep{Xu24}. The magnetic field strength often shows a slow vertical decline compared to other CGM gas phases (e.g., \citealt{Krause18,Lu23}). For example, \citet{Krause18} presented radio scale height measurements of 13 CHANG-ES galaxies (not including NGC~3556) in C- and L-bands, and a reasonable extrapolation of the synchrotron radio scale height of a normal galaxy like NGC~3556 could be $h_{\rm syn}\sim 1-1.5$~kpc. Assuming a typical synchrotron spectral index of $\alpha =1$ (comparable to the value measured between C- and L-bands of NGC~3556), the synchrotron radio scale height is linked to the magnetic scale height with the relation $h_{\rm B}=4 h_{\rm syn}$ \citep{Lu23}, so we simply assume $h_{\rm B}=5\rm~kpc$, and the magnetic field strength at the location of the background AGN SBS 1108+560 ($h\sim 30$~kpc) is $B\sim 0.02~{\rm \mu G}$. Similar as the hot gas parameter estimated above, this estimate on $B$ is just valid on order of magnitude.

\section{Initial Results and Discussions} \label{sec:ResultsDiscussions}

\subsection{Multi-phase CGM in the lower halo of NGC~3556} \label{subsec:ResultseDIG}

As shown in Fig.~\ref{fig:MultiSlitParaMapNGC3556}, the eDIG is detected typically at $\mid h\mid \lesssim(1^\prime-2^\prime)\sim(4-8)\rm~kpc$ from the midplane of NGC~3556. This vertical range is roughly comparable to the vertical extent of other CGM phases, such as the hot gas in X-ray and CR and magnetic field in radio (Fig.~\ref{fig:MultiBandImgNGC3556}). The overall H$\alpha$ and [\ion{N}{2}] line intensities, with a vignetting correction but without a flux calibration, shows an expected vertical decline (Fig.~\ref{fig:MultiSlitParaMapNGC3556}a,b). We do not present a spatial analysis of these lines at the current stage, but according to \citet{Lu23}, the H$\alpha$ scale height is $h_{\rm H\alpha}\approx 1.4\rm~kpc$, comparable to the X-ray scale height of $h_{\rm X}\approx 1.3\rm~kpc$ \citep{Li13a} and the \ion{H}{1} scale height of $h_{\rm HI}\approx 1.2\rm~kpc$ \citep{Zheng22a}. This indicates different gas phases could be spatially or even physically coupled to each other. 

Different from NGC~891 \citep{Lu24}, we do not find firm evidence of any significant global spatial variation of the [\ion{N}{2}]/H$\alpha$ line ratio (Fig.~\ref{fig:MultiSlitParaMapNGC3556}d). The origin of the apparent enhancement of the [\ion{N}{2}]/H$\alpha$ ratio on the western side is unclear, as we cannot rule out the effect of a different S/N on the eastern and western side of the galaxy caused by the vignetting (Fig.~\ref{fig:MultiSlitRawImgNGC3556}b). This probably indicates that the ionization mechanism of the eDIG does not change significantly in this moderately star forming galaxy. Although the exact ionization mechanism could be further examined with a quantitative flux calibration and observations of other emission lines, the overall [\ion{N}{2}]/H$\alpha$ line ratio of $\lesssim0.5$ and the lack of vertical variation of it strongly suggests that the eDIG is in a low ionization state with the same ionization mechanisms as the \ion{H}{2} regions in the galactic disk (e.g., \citealt{Baldwin81,Kewley13}).

A quantitative examination of the pressure balance between different CGM phases requires a careful analysis of the spatial distribution of them (e.g., \citealt{Lu23}). These will be developed in follow-up observations (deeper and covering more optical lines) and future papers. We herein only present a pilot global (instead of spatially resolved) examination of the pressure balance between a few CGM phases based on the existing data (\S\ref{subsec:OtherDataCGM}). Assuming the average magnetic field strength is $B\approx 5\rm~\mu G$ in the X-ray spectral analysis region \citep{Li13a,Xu24}, the magnetic pressure will be $P_B=\frac{B^2}{2\mu_0}\sim0.6\rm~eV~cm^{-3}$ ($\mu_0$ is the vacuum permeability), while the thermal pressure of the hot gas in the corresponding region estimated using the parameters in \S\ref{subsec:OtherDataCGM} is $P_X=n_ekT\sim0.9f_{\rm X}^{-1/2}\rm~eV~cm^{-3}$. If $f_{\rm X}\sim1$, the thermal pressure should then be comparable to the magnetic pressure. Although the density and temperature of the eDIG is not directly measured in the present paper, it is often reasonable to assume $n_e\sim0.1-1\rm~cm^{-3}$ and $T\sim10^{3-4}\rm~K$ of this relatively cool and photo-ionized gas phase at a few kpc above the disk (e.g., \citealt{Boettcher16}). In this case, the thermal pressure of the eDIG is $P_{\rm eDIG}\sim 10^{-2}-1\rm~eV~cm^{-3}$, typically no higher than that of the X-ray emitting hot gas. Therefore, the magnetic field could play an important role in the global motion of both the eDIG and the hot gas flows in the lower halo. This also explains the overall similarity of the X-ray, optical emission line, and radio continuum images (Figs.~\ref{fig:MultiSlitParaMapNGC3556}, \ref{fig:MultiBandImgNGC3556}). 

\begin{figure*}[!th]
\begin{center}
\epsfig{figure=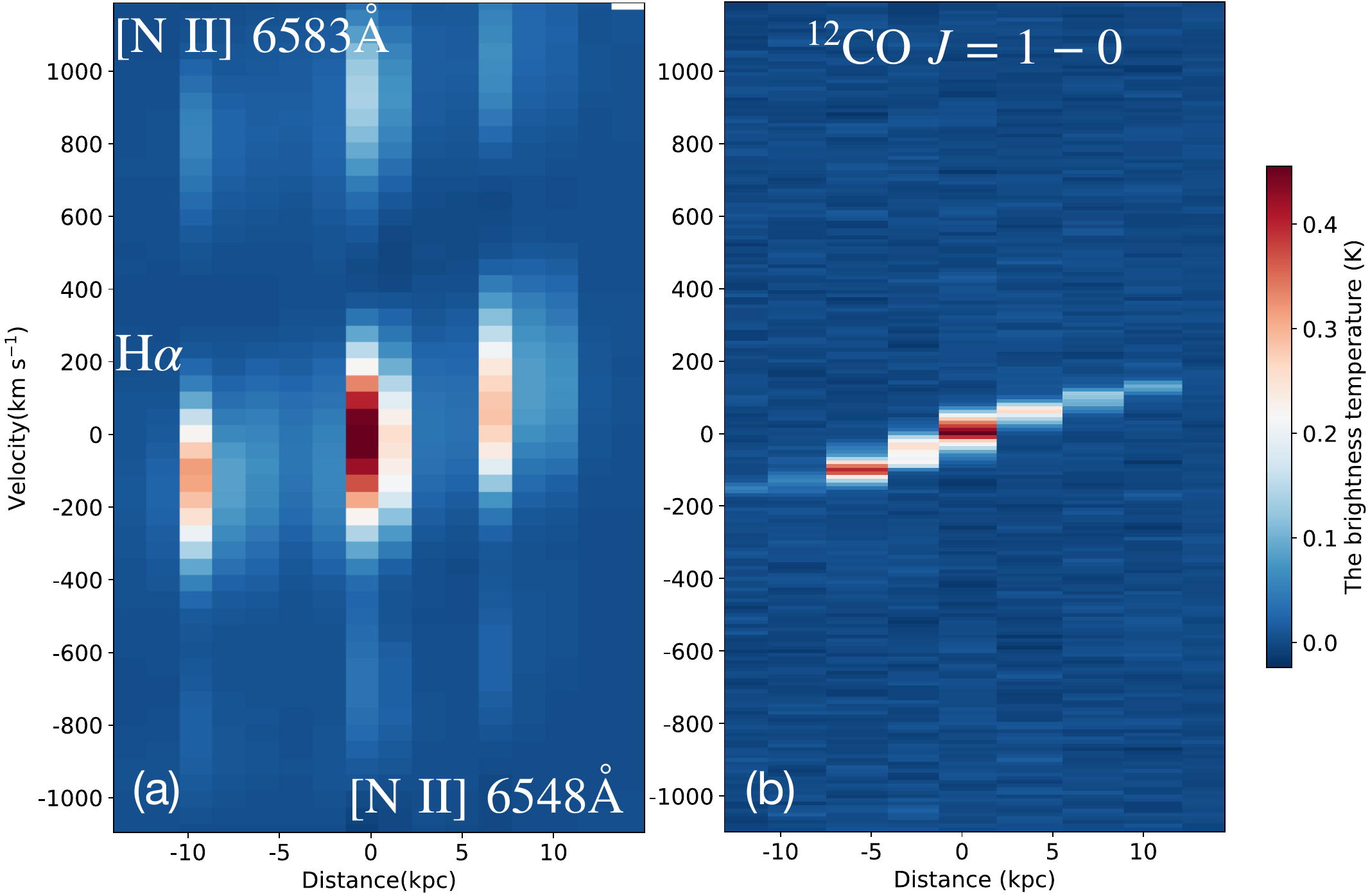,width=0.95\textwidth,angle=0}
\caption{The position-velocity (PV) diagrams along the disk of NGC~3556. The x-axis is the distance from the nucleus along the major axis, while the y-axis is the velocity from the systemic velocity of the H$\alpha$ line of the galaxy. The PV diagrams are constructed with (a) optical emission lines from our multi-slit narrow-band spectroscopy observations, with the brightness in arbitary units; and (b) the $^{12}$CO $J=1-0$ line from our IRAM 30m observations \citep{Jiang24}, with the color bar showing the brightness temperature. Locations of the multi-slit mask and the IRAM 30m beams are highlighted in Fig.~\ref{fig:MultiSlitRawImgNGC3556}. For the optical data, we bin 60 pixels ($\sim 16.4^{\prime \prime}$) along the slit direction to increase the S/N. The three parallel bands of bright features are produced by the H$\alpha$ line (in the center) and the [\ion{N}{2}] $\lambda\lambda6548,6583$~\AA\ lines (lower and upper), respectively. The pixel scales of the optical and the $^{12}$CO $J=1-0$ spectra are $\sim47$~km~s$^{-1}$ (velocity resolution $\sim190\rm~km~s^{-1}$) and $10$~km~s$^{-1}$ (also the resolution), respectively. This produces the different vertical extents of the bright features on the PV diagram. The color bar in panel~(b) shows the brightness temperature of the radio image. The slopes of the two PV diagrams are comparable to each other, indicating the two gas phases have comparable rotation velocities, and no signatures of global outflows/inflows have been revealed.
}\label{fig:PVDiagram}
\end{center}
\end{figure*}

We further compare the dynamics of the eDIG and the cold molecular gas on the galactic disk, using the multi-slit narrow-band spectroscopy data presented in this paper (Fig.~\ref{fig:MultiSlitParaMapNGC3556}c) and our IRAM 30m CO line observations of a few positions along the disk of some CHANG-ES galaxies \citep{Jiang23,Jiang24}. The slit and beam locations are plotted in Fig.~\ref{fig:MultiSlitRawImgNGC3556}a, while the results are presented as position-velocity (PV) diagrams in Fig.~\ref{fig:PVDiagram}. The PV diagram in optical band is extracted only from $\pm30\rm~pixels$ ($\pm8.19^{\prime\prime}$) along the slits above and below the midplane of the galaxy.

Apparently, the PV diagrams of the ionized gas traced by the optical emission lines and the molecular gas traced by the CO line have comparable slopes (Fig.~\ref{fig:PVDiagram}). This indicates that the dynamics of the two gas phases are both controlled by the gravity of the galaxy, and we do not see any strong signatures of global gas outflows or inflows. The corresponding rotation velocity of the galaxy is $\sim150\rm~km~s^{-1}$, comparable to those published in the archive (e.g., \citealt{Li13a}). This result also confirms that with the multi-slit narrow-band spectroscopy method, we can measure the rotation curve of NGC~3556 and other comparably massive $L^\star$ galaxies which has a maximum rotation velocity smaller than the spectral resolution of $R\sim1,600$ (or $\sim190\rm~km~s^{-1}$). The height of the bright features on the PV diagram is much larger in the optical than in radio, which is mainly caused by the different spectral resolutions. At the current spectral resolution, it is unclear if the optical lines have any intrinsic broadening. As shown in Fig.~\ref{fig:PVDiagram}a, in addition to the two bright bands (produced by [\ion{N}{2}] $\lambda\lambda6548,6583$~\AA ) up and down of the brightest band produced by H$\alpha$, there are also two bright features along the horizontal direction at the edge of the galactic disk (at $\sim -10\rm~kpc$ and $\sim +7\rm~kpc$). These features are not seen on the CO PV diagram (Fig.~\ref{fig:PVDiagram}b). Instead, except for the center, the CO emission is bright at $\sim -5\rm~kpc$. As shown in Fig.~\ref{fig:MultiSlitRawImgNGC3556}, we can see that the CO bright beam appears dark in H$\alpha$, which may be caused by the extinction of the dusty molecular clouds.

\subsection{A pilot examination of the pressure balance between the multi-phase CGM at the location of the background AGN} \label{subsec:PressureBalanceCGM}

The presence of a UV bright background AGN at such a small impact parameter of $\sim30\rm~kpc$ from a galaxy provides us with a unique opportunity to measure the column densities of different ions from the WHIM. This further helps us to examine the pressure balance between different CGM phases based on reasonable extrapolation of the multi-wavelength observations (\S\ref{subsec:OtherDataCGM}). In this section, we present a pilot comparison between the pressures of the hot gas, WHIM, and magnetic field, largely based on some very uncertain assumptions.

\begin{figure}[!th]
\begin{center}
\epsfig{figure=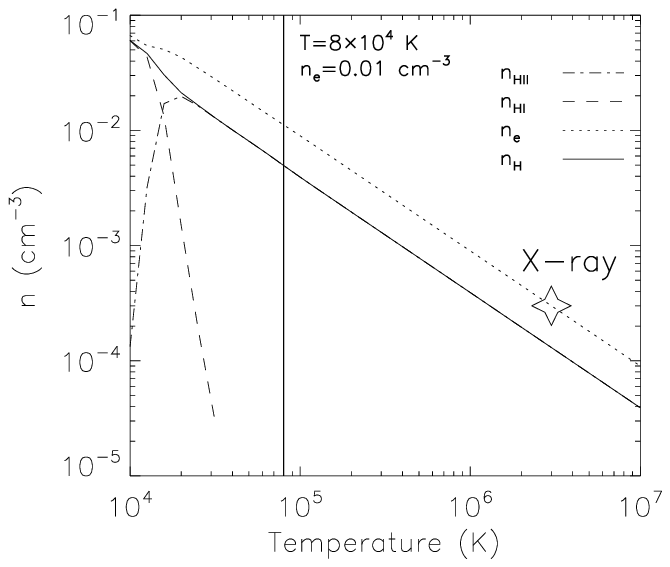,width=0.49\textwidth,angle=0, clip=}
\caption{Examination of pressure balance between the X-ray emitting hot gas and the warm/warm-hot gas traced by various UV absorption lines at the location of the background AGN SBS 1108+560. This figure shows the expected density of different particles assuming the gas is in thermal pressure equilibrium with the X-ray emitting hot gas. Measurements from the extrapolation of the \emph{Chandra} X-ray intensity profile is plotted as a star for comparison (denoted as ``X-ray''). $n_{\rm H}=n_{\rm HI}+n_{\rm HII}$ is the total hydrogen number density including both the neutral and ionized hydrogen. The thick vertical line marks where $T=8\times10^4\rm~K$ and the corresponding $n_{\rm e}$ value, which is determined in Fig.~\ref{fig:PressureBalanceXrayHighLowIons}.}\label{fig:PressureBalanceXrayUV}
\end{center}
\end{figure}

We assume there is a phase of WHIM in pressure balance with the X-ray emitting gas (using the temperature and extrapolated hot gas density from \S\ref{sec:OtherDataNGC3556}), then calculate its physical parameters. In this process, we assume the solar abundance and the ionization fraction of different elements at a given temperature from the optically thin plasma model in \citet{Bryans06}. The result is presented in Fig.~\ref{fig:PressureBalanceXrayUV}, with the curve of $n_e$ matching the X-ray measurements.

\begin{figure*}[!th]
\begin{center}
\epsfig{figure=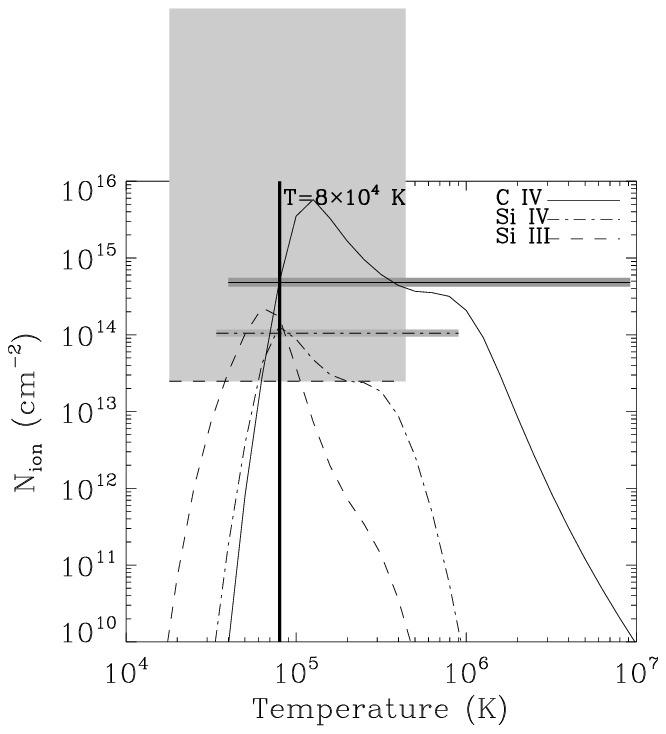,width=0.49\textwidth,angle=0, clip=}
\epsfig{figure=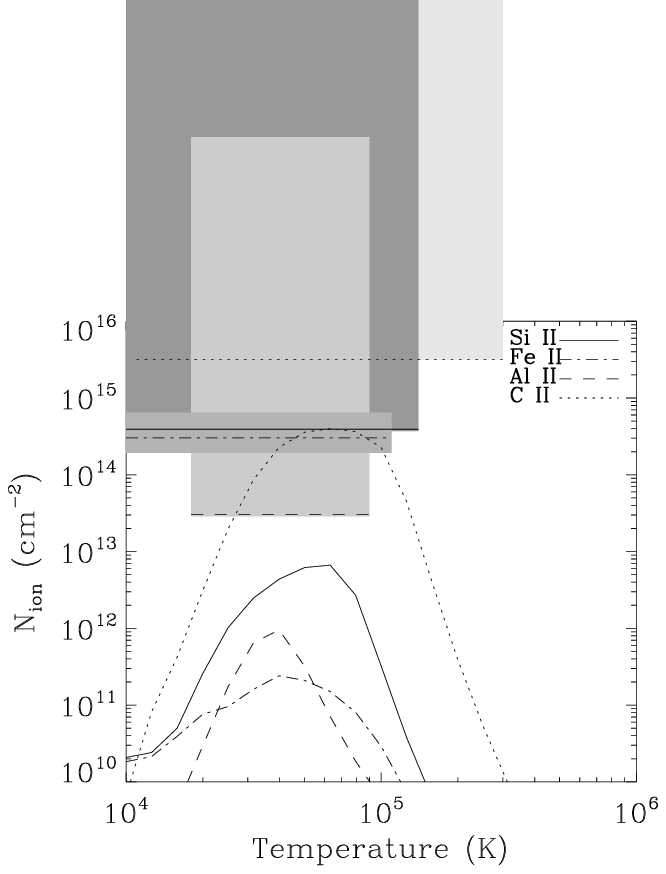,width=0.49\textwidth,angle=0, clip=}
\caption{Expected column densities of different ions in pressure balance with the X-ray emitting gas (the curves), in comparison with the observed values from the UV absorption lines (best-fit values and errors are plotted as horizontal lines and shaded area; Table~\ref{table:NionCOS}). Note that the curves are computed from the fitted lower limit on $N_{\rm HI}$, so also represent lower limits on different ions. The allow range should be above the curves. The left panel shows the high (\ion{C}{4}, \ion{Si}{4}) and intermediate (\ion{Si}{3}) ions, while the right panel shows the low ions (\ion{C}{2}, \ion{Si}{2}, \ion{Fe}{2}, and \ion{Al}{2}). The intersections between the curves and the horizontal lines are thus the expected temperatures of the gas in thermal and pressure equilibrium with hydrogen. We mark in the left panel $T=8\times10^4\rm~K$ with a thick vertical line which best matches both the \ion{C}{4} and \ion{Si}{4}, and possibly also marginally fits \ion{Si}{3} within errors. All the low ions have a measured column density far higher than the values expected from a thermal and ionization equilibrium model, indicating additional ionizing sources, such as photo-ionization. The only exception is \ion{C}{2}, which seems could be marginally described with the $T=8\times10^4\rm~K$ model. But we caution that the measurement of the \ion{C}{2} line is highly uncertain because of the saturation of the absorption (Fig.~\ref{fig:HSTCOSSpec}). Note that the $x$-axis range of (b) and (c) are different because the low ions only exist at relatively low temperatures.}\label{fig:PressureBalanceXrayHighLowIons}
\end{center}
\end{figure*}

The above estimate is the expected parameters of a solar abundance WHIM in pressure balance with the X-ray emitting gas. We next examine if this ``expected'' WHIM is consistent with the real UV absorption line measurements or not. In order to calculate the thermal pressure of the WHIM, we need to estimate its volume density from the directly measured column density of different ions (Table~\ref{table:NionCOS}). Here we assume all the ions detected in UV absorption lines have the same geometrical structures, or filling factors, i.e., $N_{\rm HI}/n_{\rm HI}=N_{\rm ion}/n_{\rm ion}$, where $N$ and $n$ are the column and volume densities of different ions. We then calculate $N_{\rm HI}/n_{\rm HI}$ under different temperatures, using the lower limit on $N_{\rm HI}$ obtained by fitting the Ly$\alpha$ line and the $n_{\rm HI}$ from Fig.~\ref{fig:PressureBalanceXrayUV} (the WHIM in pressure balance with the hot gas). We further convert $n_{\rm ion}$ in Fig.~\ref{fig:PressureBalanceXrayUV} to $N_{\rm ion}$ and compare them to the measured values Table~\ref{table:NionCOS}. Photo-ionization modeling of the UV absorption lines indicates that there are at least two different gas phases in the halo of NGC~3556: weakly ionized gas characterized with the low ions such as \ion{Si}{2}, \ion{Fe}{2}, and \ion{Al}{2}, and highly ionized gas characterized with the high ions such as \ion{C}{4} and \ion{Si}{4} (\S\ref{subsec:OtherDataHSTCOS}). We therefore show the results separately for the high and low ions in Fig.~\ref{fig:PressureBalanceXrayHighLowIons}. Note that here the curves from the ``expected'' WHIM in pressure balance with the hot CGM are calculated from the fitted lower limit on $N_{\rm HI}$, so are also lower limits on other ions.

We discuss the pressure balance between different gas phases in two cases: assuming $\log N_{\rm HI}\approx14.8$ (the lower limit in Table~\ref{table:NionCOS}) or $\log N_{\rm HI}>>14.8$. If $\log N_{\rm HI}\approx14.8$, as shown in Fig.~\ref{fig:PressureBalanceXrayHighLowIons}, all the low ions (\ion{Si}{2}, \ion{Fe}{2}, \ion{Al}{2}, including \ion{C}{2} whose column density is poorly constrained due to saturation; Fig.~\ref{fig:HSTCOSSpec}) have significantly higher column densities than expected from the pressure balance assumption. This indicates additional ionizing sources than collisional ionization under thermal equilibrium, which is likely photo-ionization as most commonly expected for low ions. On the other hand, the high ions \ion{C}{4} and \ion{Si}{4} seem well described with a thermal equilibrium model with a temperature of $T\sim8\times10^4\rm~K$ (the intermediate ion \ion{Si}{3} could also be marginally described with this model; Fig.~\ref{fig:PressureBalanceXrayHighLowIons} left). If this gas phase is in pressure balance with the hot gas, we can further estimate its electron density to be $n_{\rm e}\sim0.01\rm~cm^{-3}$ (Fig.~\ref{fig:PressureBalanceXrayUV}). On the other hand, if $\log N_{\rm HI}>>14.8$, all the curves but not the horizontal lines in Fig.~\ref{fig:PressureBalanceXrayHighLowIons} would move upward. In this case, the temperature solution of the three high and intermediate ions are unlikely significantly changed, i.e., in a relatively narrow range of $T\sim10^{4.2-5}\rm~K$. Moreover, it would be possible to find a common solution of different low ions matching the result of the high ions. It is thus not impossible that all the UV absorbing ions are in pressure balance with each other.

We further examine the role of magnetic field by calculating its pressure using the extrapolated magnetic field strength of $B\sim0.02\rm~\mu G$ at $\rho\sim30\rm~kpc$ (\S\ref{subsec:OtherDataCGM}). Since the magnetic field strength is calculated based on a CR-magnetic field energy equipartition assumption \citep{Xu24}, the CR pressure should be the same as the magnetic pressure. The corresponding magnetic pressure is $P_B=\frac{B^2}{2\mu_0}\sim10^{-5}\rm~eV~cm^{-3}$, while the thermal pressure of the hot gas or WHIM at the same location is $P_X=n_ekT\sim0.1f_{\rm X}^{-1/2}\rm~eV~cm^{-3}$, about two orders of magnitude higher than the magnetic pressure, assuming $f_{\rm X}\sim1$. Because $f_{\rm X}<1$, the magnetic pressure is in general unimportant at $\rho\sim30\rm~kpc$. This is a clearly different case than the lower halo, where the thermal and magnetic pressures are in general comparable (\S\ref{subsec:ResultseDIG}). 

The chance of finding a UV bright AGN so close to the galaxy is quite low, so the discussions presented in this section cannot be applied to most of the CHANG-ES galaxies. Moreover, the pilot examination on the pressure balance between different CGM phases at $\gtrsim10\rm~kpc$ above the disk has a lot of uncertainties caused by the extrapolation of the pressure profile of different CGM phases (\S\ref{subsec:OtherDataCGM}). For example, the adopted synchrotron emission scale height is calculated without subtracting the thermal contribution in radio near the galactic disk \citep{Krause18}. This means the actual radio scale height may be larger. Furthermore, \citet{Pakmor17} shows in MHD simulations that the radio profile flattens at large vertical heights. This is also consistent with the latest detection of large-scale radio features with the low-frequency LOFAR observations \citep{Heald22,Heesen23}. If the non-thermal radio profile is indeed flatter than we have assumed, the magnetic pressure could be significantly higher. On the other hand, the X-ray intensity profile is also poorly constrained for moderately star forming $L^\star$ galaxies such as NGC~3556. We have adopted the profile for massive spiral galaxies, which in principle could be richer in hot gas, but are actually X-ray faint \citep{Li17,Li18}. It is not impossible that NGC~3556 is even X-ray fainter than what we have expected from the extrapolated X-ray intensity profile of these massive galaxies. Considering the bias in both radio and X-ray measurements, it is not impossible that the hot gas and magnetic field may be closer to pressure balance than we have examined above.

\section{Summary} \label{sec:Summary}

In this paper, we introduce a novel multi-slit narrow-band spectroscopy method to conduct spatially resolved spectroscopy observations of extended optical emission line objects. We made five slit masks, each with 40 parallel longslits uniformly distributed in a $\sim18.5^\prime$ FOV, to be used together with the OSMOS imager/spectrograph on the MDM 2.4m telescope and any narrow-band filters with a bandwidth $\lesssim200$\AA . When all of the five masks are used (not simultaneously), the total sky coverage will be $\sim23\%$ while the angular resolution is $5^{\prime\prime}$/seeing limited perpendicular/along the slits, reaching an efficiency comparable to a large FOV IFU.

We apply this new multi-slit narrow-band spectroscopy method to the eDIG-CHANGES project, which is a systematic study of the eDIG around a sample of nearby edge-on disk galaxies (the CHANG-ES sample). We briefly introduce the overall project design and major scientific goals of the eDIG-CHANGES project. We further present an initial pilot case study of NGC~3556, a galaxy with moderate properties within the sample, while detailed data reduction procedures will be introduced in \citet{Lu24}. The eDIG is detected to a vertical extent of $\sim4-8\rm~kpc$ above the disk, comparable to our X-ray and radio observations. We do not see any significant trend of the vertical variation of the [\ion{N}{2}]/H$\alpha$ line ratio, which is also low and consistent with those expected from disk \ion{H}{2} regions. A rough examination of the pressure balance between different CGM phases indicates that the magnetic field may play an important role in the global motion of both the eDIG and the hot gas in the lower halo within a few kpc from the galactic disk. We also study the dynamics of the multi-phase gases by comparing the PV diagrams of the ionized gas traced by our optical emission line observations and that of the molecular gas traced by our IRAM 30m CO~$J=1-0$ observations. Dynamics of the two gas phases are both controlled by the gravity of the galaxy, so they show similar rotation curves with a maximum rotation velocity of $\sim150\rm~km~s^{-1}$, below our optical spectral resolution but well constrained with the existing multi-slit narrow-band spectra. We do not find any significant global outflow/inflow features. The lack of global motions of the gas may be consistent with the pressure balance close to the galactic disk, which means different CGM phases all move together. This scenario, however, needs to be further examined with deeper multi-wavelength data.

There is also a UV bright background AGN projected at an impact parameter of only $\approx29\rm~kpc$ from the center of NGC~3556. Using the archival HST/COS data, we detected many UV absorption lines and conducted Cloudy modeling of them. The UV-absorbing gas can be modeled with at least two gas phases, a higher temperature gas phase traced by some high and intermediate ions and a lower temperature gas phase traced by only the low ions. Extrapolating the X-ray and radio data to the location of this background AGN indicates that the high ions could be in pressure balance with the X-ray emitting hot gas, while the condition of the low ions is quite uncertain, caused by the saturation of some UV absorption lines. Different from the lower halos, the magnetic pressure is not important at large vertical distances from the galactic disk, although the extrapolation of the pressure profiles to larger heights may cause some biases.

\acknowledgments

The authors acknowledge the late R. Mark Wagner from The Ohio State University, who helped us make the multi-slit masks. We also acknowledge Eric Galayda and Justin Rupert at the MDM observatory for the help in the multi-slit spectroscopy observations, and Dr. Rainer Beck from the Max-Planck-Institut f\"{u}r Radioastronomie for scientific discussions. J.T.L. acknowledges the financial support from the National Science Foundation of China (NSFC) through the grant 12273111. T.F. is supported by the National Key R\&D Program of China under No. 2017YFA0402600, the National Natural Science Foundation of China under Nos. 11890692, 12133008, 12221003, and the science research grant from the China Manned Space Project with No. CMS-CSST-2021-A04. H.L. is supported by the National Key R\&D Program of China No. 2023YFB3002502 and the National Natural Science Foundation of China under No. 12373006. Research in this field at AIRUB is supported by Deutsche Forschungsgemeinschaft SFB 1491. Y.Y. acknowledges support from the National Natural Science Foundation of China (NSFC) through the grant 12203098. T.W. acknowledges financial support from the grant CEX2021-001131-S funded by MCIU/AEI/ 10.13039/501100011033, from the coordination of the participation in SKA-SPAIN, funded by the Ministry of Science, Innovation and Universities (MCIU).

\bibliography{eDIGCHANGES}

\end{CJK*}
\end{document}